\documentclass[twocolumn,english]{article}
\usepackage[T1]{fontenc}
\usepackage[latin9]{inputenc}
\usepackage[a4paper]{geometry}
\usepackage{babel}
\usepackage{bm}
\usepackage{amsmath}
\usepackage{amssymb}
\usepackage{graphicx}
\usepackage[unicode=true]
 {hyperref}

\makeatletter
\usepackage{babel}
\usepackage{xcolor}
\definecolor{myred}{rgb}{0.66, 0.15, 0.15}
\definecolor{darkgreen}{rgb}{0.0, 0.5, 0.0}

\usepackage{hyperref}          
\hypersetup{
     colorlinks   = true,
     citecolor    = myred,
     urlcolor     = myred,
     urlbordercolor = myred
}

\usepackage[normalem]{ulem}
\usepackage{flushend}
\usepackage{cuted}

\usepackage{cite}

\makeatother

\begin{document}
\begin{strip}

{\Large{}\vspace{-1cm}
}\textsf{\textbf{\huge{}Universal geometry of two-neutron halos and
Borromean Efimov states close to dissociation\vspace{0.5cm}
}}{\huge\par}

{\Large{}Pascal Naidon}\textsf{\textbf{\huge{}\vspace{0.2cm}
}}{\huge\par}

{\small{}Few-Body Systems Physics Laboratory, RIKEN Nishina Centre,
RIKEN, Wak{\={o}}, 351-0198 Japan.}\textit{ }{\Large{}\vspace{-0.4cm}
}{\Large\par}

\textit{\href{mailto:pascal@riken.jp}{pascal@riken.jp}}\textsf{\textbf{\huge{}\vspace{0.2cm}
}}{\huge\par}

\today{\Large{}\vspace{0cm}
}{\Large\par}
\begin{abstract}
The geometry of Borromean three-body halos, such as two-neutron halo
nuclei or triatomic molecules close to dissociation, is investigated
using a three-body model. This model enables to analytically derive
the universal geometric properties found recently within an effective-field
theory for halos made of a core and two resonantly-interacting particles
{[}Phys. Rev. Lett., 128, 212501 (2022){]}. It is shown that these
properties not only apply to the ground three-body state, but also
to all the excited (Efimov) states where the core-particle interaction
is resonant. Furthermore, a universal geometry persists away from
the resonant regime between the two particles, for any state close
to the three-body threshold. This ``halo universality'', which applies
equally to all states, is different from the Efimov universality,
which is only approximate for the ground state. It is explained by
the separability of the hyper-radius and hyper-angles close to the
three-body dissociation threshold.
\end{abstract}
\end{strip}

\section{Introduction\label{sec:Introduction}}

\hypersetup{linkcolor=myred}

Quantum halos~\cite{Jensen2004}, i.e. quantum few-body bound states
whose spatial extent exceeds the range of the bodies' interactions,
have been studied for over four decades since the experimental discovery
of halos in atomic nuclei in the 1980s~\cite{Tanihata1985,Tanihata1985a,AlKhalili2004,Tanihata2013},
followed by the controlled creation of halos in ultracold-atom experiments
from the 2000s~\cite{Kraemer2006,Braaten2007,Zirbel2008,Chin2010}.
Quantum halo systems can be composed of identical particles loosely
bound to each other, or as is often the case for halo nuclei, a composite
core and a few loosely bound particles forming the halo. Moroever,
quantum halos can be Borromean~\cite{Zhukov1993}, i.e. they do not
remain bound if one of the particles is removed. The geometry of
quantum halos is characterised by large mean square radii~\cite{Fedorov1994a,Yamashita2004,Bertulani2007,Hagino2007,Canham2008,Yamashita2011a,Acharya2013},
which can be extracted from experimental measurements~\cite{Marques2000,Nakamura2006,Tanaka2010,Moriguchi2013,Mosby2013,Togano2016}.
In a recent work based on an effective-field theory~\cite{Hongo2022},
universal analytical relations were found between different mean square
radii for Borromean three-body halos made of a core and two resonantly-interacting
particles, such as two-neutron halo nuclei.

In the present work, it is shown how the analytic relations can be
obtained from the Faddeev approach to the three-body problem. Although
it is implicit in the work of Ref.~ \cite{Hongo2022}, it is here
emphasised that the analytical relations apply not only to a ground
state but also to excited Borromean halo states. The applicability
of the relations is then tested numerically within a separable three-body
model where the three-body Efimov effect occurs. These numerical calculations
confirm that the analytic properties are relevant to any Borromean
halo state, i.e. any Efimov state close to the three-body dissociation.
Finally, the limiting values close and away from the two-particle
resonance are retrieved analytically in the hyper-spherical representation.

\section{Model\label{sec:Model}}

\begin{figure}
\begin{centering}
\includegraphics[width=7cm]{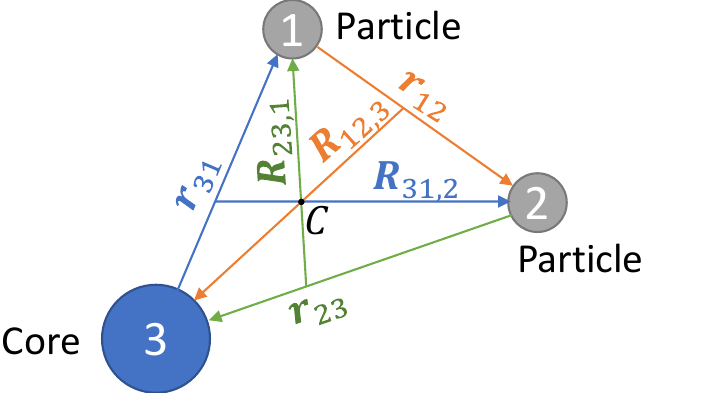}
\par\end{centering}
\caption{\label{fig:Schematic-representation}Schematic representation of a
system formed by a core (3) and two identical particles (1 and 2).}

\end{figure}
We start with a three-body model for a core particle denoted by 3,
of mass $m_{3}$, interacting with two identical particles denoted
by 1 and 2, of mass $m_{1}=m_{2}$ -- see Fig.~\ref{fig:Schematic-representation}.
The corresponding Schrödinger equation for the three-body wave function
$\tilde{\Psi}$ in momentum representation reads
\begin{equation}
\left(\sum_{i=1}^{3}\frac{\hbar^{2}k_{i}^{2}}{2m_{i}}+\sum_{i<j}\hat{V}_{ij}-E\right)\tilde{\Psi}=0,\label{eq:SchrodingerEquation}
\end{equation}
where $E$ is the total energy and $\hat{V}_{ij}$ are the respective
two-body interaction operators, having ranges denoted as $\Lambda_{ij}^{-1}$.
In the present notation, an operator $\hat{O}_{ij}$ acts on the relative
wave vector $\bm{k}_{ij}=\frac{m_{i}\bm{k}_{j}-m_{j}\bm{k}_{i}}{m_{i}+m_{j}}$
between particles $i$ and $j$. Introducing the Faddeev components~\cite{Faddeev1961}
$\mathcal{F}_{ij}=\hat{V}_{ij}\tilde{\Psi}$ , one can write for $E<0$,
\begin{equation}
\tilde{\Psi}=\frac{\mathcal{F}_{12}+\mathcal{F}_{23}+\mathcal{F}_{31}}{E-\sum_{i=1}^{3}\frac{\hbar^{2}k_{i}^{2}}{2m_{i}}}.\label{eq:ThreeBodyWavefunction}
\end{equation}
It follows from the definition of the Faddeev components that they
satisfy the following Faddeev equations:
\begin{align}
\mathcal{F}_{ij} & =\hat{T}_{ij}\left(z_{ij}\right)\frac{\mathcal{F}_{jk}+\mathcal{F}_{ki}}{E-\sum_{i=1}^{3}\frac{\hbar^{2}k_{i}^{2}}{2m_{i}}},\label{eq:FaddeevEquations}
\end{align}
where the two-body transition operators $\hat{T}_{ij}(z)$ are defined
from the original interactions $\hat{V}_{ij}$ by:
\begin{equation}
\hat{T}_{ij}(z)=\hat{V}_{ij}+\left(\hat{V}_{ij}G_{ij}^{+}(z)\right)\hat{T}_{ij}(z),\label{eq:TwoBodyTransitionOperator}
\end{equation}
where $G_{ij}^{+}(z)=\left(z+i0^{+}-\frac{\hbar^{2}}{2\mu_{ij}}k_{ij}^{2}\right)^{-1}$
and $\mu_{ij}=\left(\frac{1}{m_{i}}+\frac{1}{m_{j}}\right)^{-1}$
is the reduced mass for particles $i$ and $j$. The two-body energy
$z_{ij}$ in Eq.~(\ref{eq:FaddeevEquations}) is obtained by subtracting
from the total energy the centre-of-mass kinetic energy and the relative
kinetic energy between the dimer $(i,j)$ and particle $k$, 
\begin{equation}
z_{ij}=E-\frac{\hbar^{2}}{2(m_{1}+m_{2}+m_{3})}K_{C}^{2}-\frac{\hbar^{2}}{2\mu_{ij,k}}K_{ij,k}^{2},\label{eq:TwoBodyEnergy}
\end{equation}
where $\bm{K}_{C}=\bm{k}_{1}+\bm{k}_{2}+\bm{k}_{3}$ is the total
wave vector, $\mu_{ij,k}=\left(\frac{1}{m_{i}+m_{j}}+\frac{1}{m_{k}}\right)^{-1}$
is the reduced mass and $\bm{K}_{ij,k}=\frac{(m_{i}+m_{j})\bm{k}_{k}-m_{k}(\bm{k}_{i}+\bm{k}_{j})}{m_{i}+m_{j}+m_{k}}$
is the relative wave vector between the dimer $(i,j)$ and particle
$k$. Since the system is translationally invariant, it will be assumed
in the following that the total wave vector $\bm{K}_{C}$ is zero.

\section{Mean Square Radii\label{sec:Mean-Square-Radii}}

Following Ref.~\cite{Hongo2022}, let us denote the mass ratio $m_{3}/m_{1}$
as $A$, and define the following mean square \emph{matter radius}
$\langle r_{m}^{2}\rangle$ and mean square \emph{core radius} $\langle r_{c}^{2}\rangle$
(or \emph{charge radius}, if the core carries an electric charge)
\begin{align}
\langle r_{m}^{2}\rangle & \equiv\frac{2\langle r_{C1}^{2}\rangle+A\langle r_{C3}^{2}\rangle}{A+2}\label{eq:SquareMatterRadius}\\
\langle r_{c}^{2}\rangle & \equiv\langle r_{C3}^{2}\rangle\label{eq:SquareChargeRadius}
\end{align}
where $C$ denotes the centre of mass of the three particles, and
$\bm{r}_{Ci}=\bm{r}_{i}-\bm{r}_{C}$ is the relative vector between
particle $i$ and the centre of mass $C$. One can easily show that
\begin{align}
r_{C1} & =\frac{A+1}{A+2}R_{23,1}\label{eq:rC1}\\
r_{C3} & =\frac{2}{A+2}R_{12,3}\label{eq:rC3}
\end{align}
where $R_{ij,k}$ is the relative vector between particle $k$ and
the centre of mass of the dimer $(i,j)$ - see Fig.~\ref{fig:Schematic-representation}.
It follows that
\begin{equation}
\frac{\langle r_{m}^{2}\rangle}{\langle r_{c}^{2}\rangle}=\frac{\frac{1}{2}\left(A+1\right)^{2}\frac{\langle R_{23,1}^{2}\rangle}{\langle R_{12,3}^{2}\rangle}+A}{A+2}.\label{eq:SquareRadiusRatio}
\end{equation}
The two averages $\langle R_{23,1}^{2}\rangle$ and $\langle R_{12,3}^{2}\rangle$
may be calculated from the wave function $\tilde{\Psi}$ in momentum
representation as
\begin{align}
\langle R_{23,1}^{2}\rangle & =\frac{\int d^{3}K_{23,1}d^{3}k_{23}\left|\bm{\nabla}_{\bm{K}_{23,1}}\tilde{\Psi}\right|^{2}}{\int d^{3}K_{23,1}d^{3}k_{23}\left|\tilde{\Psi}\right|^{2}},\label{eq:SquareRadius32-1}\\
\langle R_{12,3}^{2}\rangle & =\frac{\int d^{3}K_{12,3}d^{3}k_{12}\left|\bm{\nabla}_{\bm{K}_{12,3}}\tilde{\Psi}\right|^{2}}{\int d^{3}K_{12,3}d^{3}k_{12}\left|\tilde{\Psi}\right|^{2}}.\label{eq:SquareRadius12-3}
\end{align}

\section{Two-particle resonance\label{sec:Two-particle-resonance}}

Let us now consider a bound core-particle-particle system (in a ground
or excited state) close to the three-body dissociation threshold $E\to0^{-}$,
so that it becomes a halo whose extent exceeds the ranges $\Lambda_{ij}^{-1}$
of the particles' interactions. In this situation, the calculation
of the square radii is dominated by the low-momentum part of the wave
function, $k,K\ll\Lambda_{ij}$. At low momenta and energy, the (on-
and off-shell) two-body T-matrix elements are given by\footnote{See Appendix~\ref{sec:Low-momentum-and-energy} for details.}
\begin{equation}
\langle\bm{k}\vert\hat{T}_{ij}(z_{ij})\vert\bm{k}^{\prime}\rangle\approx\frac{4\pi\hbar^{2}}{2\mu_{ij}}\left(\frac{1}{a_{ij}}+i\sqrt{\frac{2\mu_{ij}}{\hbar^{2}}z_{ij}}\right)^{-1},\label{eq:Off-shell-value}
\end{equation}
where $a_{ij}$ is the s-wave scattering length between particles
$i$ and $j$. It follows from Eq.~(\ref{eq:FaddeevEquations})
that each Faddeev component $\mathcal{F}_{ij}$ at low momenta is
proportional to the right-hand side of Eq.~(\ref{eq:Off-shell-value}).

Close to a resonance between particles 1 and 2, such that $\vert a_{12}^{-1}\vert\ll\vert a_{23}^{-1}\vert,\vert a_{31}^{-1}\vert,\Lambda_{ij}$,
and for sufficiently small energy $\sqrt{2\mu_{12}\vert E\vert}/\hbar\ll\vert a_{23}^{-1}\vert,\vert a_{31}^{-1}\vert$
of the three-body system, the Faddeev component $\mathcal{F}_{12}$
at low momenta $k_{12}\ll\Lambda_{12}$ and $K_{12,3}\lesssim\sqrt{2\mu_{12,3}\vert E\vert}/\hbar\ll\Lambda_{12}$
becomes dominant over $\mathcal{F}_{23}$ and $\mathcal{F}_{31}$
in the three-body wave function of Eq.~(\ref{eq:ThreeBodyWavefunction}).
Furthermore, $\mathcal{F}_{12}$ is proportional to $\langle\bm{0}\vert\hat{T}_{12}\vert\bm{0}\rangle$
and depends only on $K_{12,3}$:
\begin{equation}
\mathcal{F}_{12}\propto\mathcal{F}(K_{12,3})=\frac{4\pi\hbar^{2}/2\mu_{12}}{\frac{1}{a_{12}}-\sqrt{-\frac{2\mu_{12}}{\hbar^{2}}E+\frac{\mu_{12}}{\mu_{12,3}}K_{12,3}^{2}}}.\label{eq:F12approximate}
\end{equation}
The calculation of the square radii is thus simplified and yields
\begin{equation}
\frac{\langle r_{m}^{2}\rangle}{\langle r_{c}^{2}\rangle}=\frac{A}{2}\left(1+\frac{f_{n}}{f_{c}}\right)\label{eq:SquareRadiusRatioTer}
\end{equation}
where the quantities $f_{n}$ and $f_{c}$ are determined solely by
the Faddeev component $\mathcal{F}$ associated with the two resonant
particles, 
\begin{align}
f_{n} & \equiv\frac{1}{2}\int d^{3}K\frac{\mathcal{F}(K)^{2}}{\tilde{K}^{3}}\label{eq:f1}\\
f_{c} & \equiv\int d^{3}K\left(\frac{\mathcal{F}^{\prime}(K)^{2}}{\tilde{K}}-\frac{K\mathcal{F}(K)\mathcal{F}^{\prime}(K)}{\tilde{K}^{3}}+\frac{K^{2}\mathcal{F}(K)^{2}}{2\tilde{K}^{5}}\right)\label{eq:f2}
\end{align}
with $\tilde{K}^{2}=K^{2}-\frac{2\mu_{12,3}E}{\hbar^{2}}$. For
$a_{12}<0$ and $E<0$, it can be shown\footnote{See Appendix \ref{sec:Functions} for details.}
that $f_{n}$ and $f_{c}$ reduce to the following integrals
\begin{align}
f_{n}(\beta) & \propto\int_{1}^{\infty}dy\frac{\sqrt{y-1}}{2y^{3/2}\left(\beta+\sqrt{y}\right)^{2}},\label{eq:f1simplified}\\
f_{c}(\beta) & \propto\int_{1}^{\infty}\frac{dy}{2\sqrt{y(y-1)}\left(\beta+\sqrt{y}\right)^{2}},\label{eq:f2simplified}
\end{align}
where 
\begin{equation}
\beta=\sqrt{\frac{E_{12}}{\vert E\vert}}\label{eq:beta}
\end{equation}
 is the square root ratio of the particle-particle virtual energy
$E_{12}=\frac{\hbar^{2}}{2\mu_{12}\vert a_{12}\vert^{2}}$ and the
trimer binding energy $\vert E\vert$. The above integrals were shown
in Ref.~\cite{Hongo2022} to admit the following analytical expressions,
\begin{align}
f_{n}(\beta) & \propto\begin{cases}
{\displaystyle \beta^{-3}\left(\pi-2\beta+(\beta^{2}-2)\frac{\arccos\beta}{\sqrt{1-\beta^{2}}}\right),} & \beta<1\\
\\
{\displaystyle \beta^{-3}\left(\pi-2\beta+(\beta^{2}-2)\frac{\text{arccosh}\beta}{\sqrt{\beta^{2}-1}}\right),} & \beta>1
\end{cases}\label{eq:f1explicit}\\
f_{c}(\beta) & \propto\begin{cases}
{\displaystyle \frac{1}{1-\beta^{2}}-\frac{\beta\arccos\beta}{(1-\beta^{2})^{3/2}},} & \beta<1\\
\\
{\displaystyle -\frac{1}{\beta^{2}-1}+\frac{\beta\text{arccosh}\beta}{(\beta^{2}-1)^{3/2}},} & \beta>1
\end{cases}\label{eq:f2explicit}
\end{align}

Thus, the ratio of matter and charge radii of Eq.~(\ref{eq:SquareRadiusRatioTer})
is universally determined by the mass ratio $A$ and the square root
ratio $\beta$. In particular, it tends to $A$ for $\beta\gg1$
and to $\frac{2}{3}A$ for $\beta\ll1$. More generally, any geometric
property of the system that does not depend explicitly on its size,
such as length ratios and angles, is universally determined by $A$
and $\beta$, owing to the form of Eq.~(\ref{eq:F12approximate})
and the fact that $\mathcal{F}_{23},\mathcal{F}_{31}\ll\mathcal{F}_{12}$.
The word \emph{universal} here means independent of the details of
the interactions, but it also means independent on whether the considered
state is excited or not, as we shall confirm numerically below.

\section{Numerical investigation\label{sec:Numerical-investigation}}

In order to test the validity of the previous analytical results,
the system is now investigated numerically close to and far from the
two-body resonance, for states of total angular momentum equal to
zero. The two particles are assumed to be in a symmetric state, i.e.
they correspond to either two identical bosons, or two identical fermions
in antisymmetric spin states, such as two neutrons in a singlet state.
For simplicity, the interactions are taken to be separable, $\langle\bm{k}\vert\hat{V}_{ij}\vert\bm{k}^{\prime}\rangle=g_{ij}\phi_{ij}(\bm{k})\phi_{ij}(\bm{k}^{\prime})$.
In this case, the transition operators $\hat{T}_{ij}$ are also separable,
and the Faddeev Eqs.~(\ref{eq:FaddeevEquations}) simplify to integral
equations of the Skorniakov$-$Ter-Martirosian (STM) type~\cite{Skorniakov1957},
which can easily be solved numerically. The form factors $\phi_{ij}$
are chosen to be of the Gaussian type, $\phi_{ij}(\bm{k})=\exp(-k^{2}/\Lambda_{ij}^{2})$,
where $\Lambda_{ij}^{-1}$ characterise the range of interactions.
The interactions being of the same physical nature, $\Lambda_{ij}$
are taken for simplicity to be all equal to the same order of magnitude
$\Lambda$. The strengths $g_{12}$ and $g_{23}=g_{31}$ determine
the respective scattering lengths $a_{12}$ and $a_{23}=a_{31}$.

\begin{figure}
\includegraphics[viewport=0bp 28bp 270bp 156bp,clip,width=8.5cm]{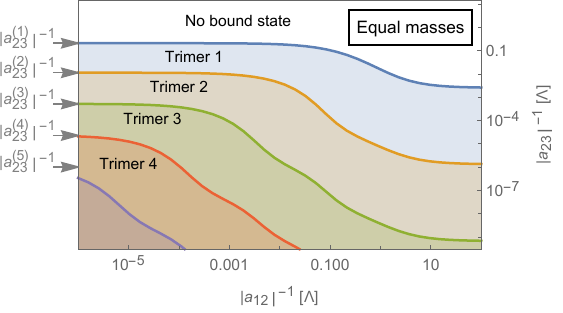}

\includegraphics[width=8.5cm]{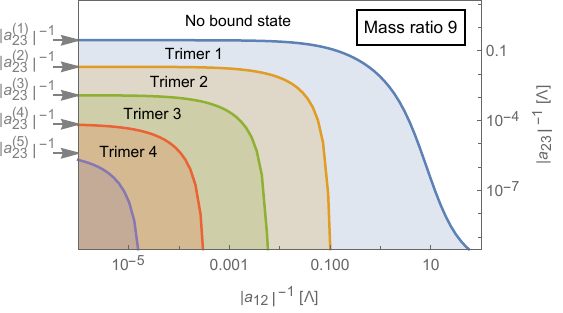}

\caption{\label{fig:Regions-of-existence}Regions of existence of Borromean
bound states of a core and two identical particles, as a function
of the core-particle inverse scattering length $\vert a_{23}\vert^{-1}$
and particle-particle inverse scattering length $\vert a_{12}\vert^{-1}$.
Top panel: equal mass case $A=1$. Bottom panel: heavy core and light
particle, with mass ratio $A=9$. The arrows show the critical core-particle
scattering lengths $a_{23}^{(n)}$ to get an $n$th bound state.}

\end{figure}

Figure~\ref{fig:Regions-of-existence} shows the regions of existence
of three-body bound states as a function of $1/a_{12}<0$ and $1/a_{23}<0$,
i.e. in the Borromean region where there are no two-body bound states.
One can see that the core-particle scattering length $\vert a_{23}\vert$
must exceed a critical value $\vert a_{23}^{(1)}\vert$ in order to
allow a first three-body bound state. Owing to the Efimov effect~\cite{Efimov1970a,Amorim1997,Braaten2006,Canham2008,Naidon2017},
there is an infinite number of three-body bound states as $\vert a_{23}\vert$
is further increased towards infinity. The curves in Fig.~\ref{fig:Regions-of-existence}
represent the sucessive thresholds for the appearance of these trimer
states. Conversely, these curves can also be regarded as the thresholds
where the trimer states dissociate into three unbound particles when
the interactions are weakened. It is near these thresholds that the
trimer becomes a large halo. Due to the Efimov effect, the thresholds
follow a geometric progression in the limit of highly-excited states.
However, for large mass ratio $A$ and small scattering length $\vert a_{12}\vert$,
the discrete scaling factor is so large that only the ground state
is observable, while the excited states are too weakly bound to be
seen. We thus focus on the region of large scattering length $\vert a_{12}\vert$
relevant to two identical particles close to unitarity such as two
neutrons.

\begin{figure}
\includegraphics[width=7cm]{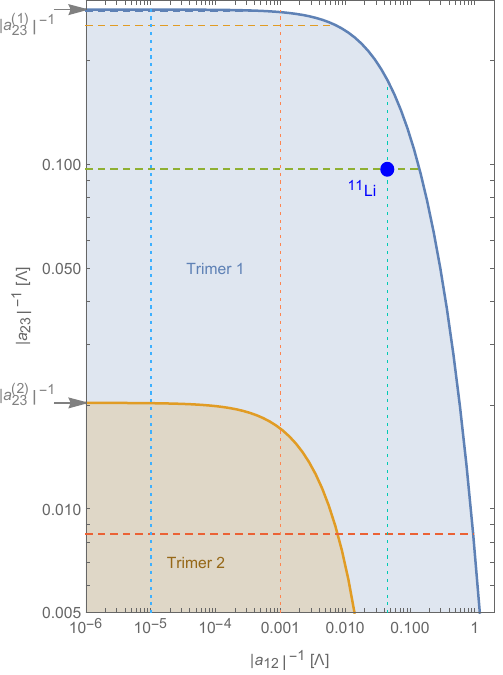}

\caption{\label{fig:Close-up-figure}Close-up figure of the bottom panel of
Fig.~\ref{fig:Regions-of-existence}. The horizontal dashed lines
correspond to the values $\vert a_{23}\vert^{-1}=0.99,0.9,0.344,0.03$
in units of $\vert a_{23}^{(1)}\vert^{-1}$. The vertical dotted lines
correspond to the values $\vert a_{23}\vert^{-1}=10^{-5},10^{-3},0.0447$
in units of $\Lambda$.}

\end{figure}

\begin{figure}
\includegraphics[viewport=0bp 27bp 285bp 191bp,clip,width=8cm]{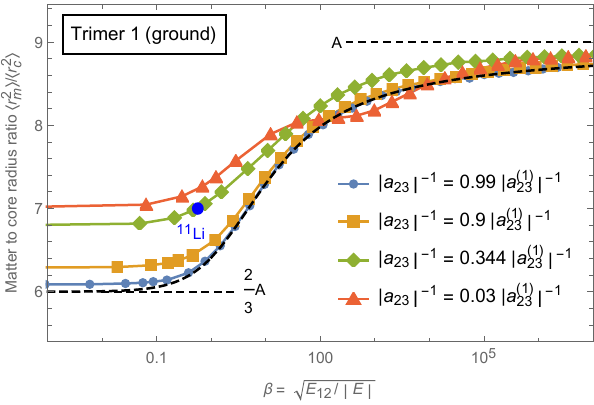}

\includegraphics[viewport=0bp 27bp 285bp 191bp,clip,width=8cm]{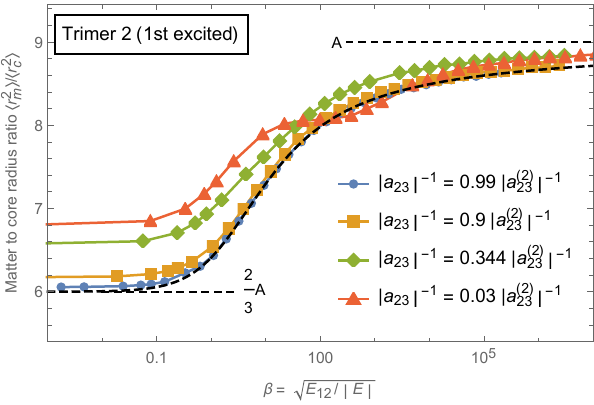}

\includegraphics[width=8cm]{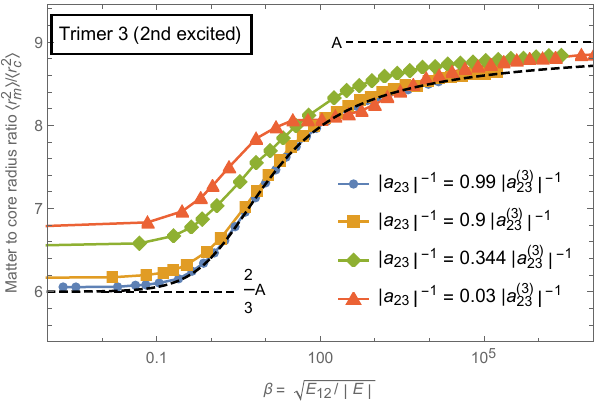}

\caption{\label{fig:Ratio}Matter/Core radius ratio $\langle r_{m}^{2}\rangle/\langle r_{c}^{2}\rangle$
as a function of the square root ratio $\beta$ of the two-particle
virtual binding energy $E_{12}$ and the trimer binding energy $\vert E\vert$.
The different curves correspond to different values of $a_{23}$,
which for the ground-state trimer are shown by horizontal dashed lines
in Fig.~\ref{fig:Close-up-figure}. The dashed curve corresponds
to the analytical formula given by Eqs.~(\ref{eq:SquareRadiusRatioTer},~\ref{eq:f1explicit},~\ref{eq:f2explicit})
and the horizontal dotted lines show the limits $\frac{2}{3}A$ and
$A$. Top panel: ground-state trimer ; middle panel: first excited
trimer. Bottom panel: second excited trimer.}
\end{figure}

This region is shown for $A=9$ in Fig.~\ref{fig:Close-up-figure},
where the ground-state and first-excited state are visible. Let us
first scan the ground state by varying the particle-particle scattering
length $a_{12}$ as indicated by the horizontal dashed lines, and
calculate its geometric properties such as the matter/core radius
ratio $\langle r_{m}^{2}\rangle/\langle r_{c}^{2}\rangle$ using Eqs.~(\ref{eq:SquareRadius32-1}-\ref{eq:SquareRadius12-3}).
The result is shown in the top panel of Fig.~\ref{fig:Ratio}. The
analytical formula based on Eqs.~(\ref{eq:SquareRadiusRatioTer},~\ref{eq:f1explicit},~\ref{eq:f2explicit})
found in Ref.~\cite{Hongo2022} accurately reproduces the numerical
calculations when $\vert a_{23}\vert$ is within $1\%$ of the critical
value $\vert a_{23}^{(1)}\vert$. However, there are significant deviations
for $\vert a_{23}\vert\gtrsim\vert a_{23}^{(1)}\vert$. Therefore,
 it appears that the analytical formula does require a fine tuning
of the core-particle interaction. Nevertheless, for values of $\vert a_{12}\vert$
very close to the threshold, the ratio $\langle r_{m}^{2}\rangle/\langle r_{c}^{2}\rangle$
always tends to the limit $A$, regardless of the core-particle scattering
length $a_{23}$.

The same situation is observed for excited states, as shown in the
middle and bottom panels of Fig.~\ref{fig:Ratio}. The results for
the first two excited states look nearly identical, which is expected
since they follow a discrete scaling invariance associated with the
Efimov effect, whereas small differences can be seen for the ground
state, which is also expected since it deviates more strongly from
the discrete scaling invariance. However, the differences remain small,
and the results are identical for all states close to the dissociation
threshold $E\to0^{-}$, i.e. either $\vert a_{23}\vert\approx\vert a_{23}^{(i)}\vert$
or $\beta\gg1$.

\section{Hyper-spherical representation\label{sec:Hyper-spherical-representation}}

The universality of the geometry close to the dissociation threshold
can be understood using the hyper-spherical coordinates constituted
by the hyper-radius $R=\sqrt{x_{k}^{2}+y_{k}^{2}}$ giving the global
size of the trimer and the hyper-angles, such as $\alpha_{k}=\arctan\frac{y_{k}}{x_{k}}$,
describing its shape, where $\bm{x}_{k}=\left(\mu_{ij}/m\right)^{1/2}\bm{r}_{ij}$,
$\bm{y}_{k}=\left(\mu_{ij,k}/m\right)^{1/2}\bm{R}_{ij,k}$, and $m$
is a normalisation mass which can be taken to be the particles' mass
$m_{1}=m_{2}$. In these coordinates, the wave function $\Psi$ of
a halo state with zero total angular momentum admits the following
hyperspherical adiabatic expansion~\cite{Nielsen2001},
\begin{equation}
\Psi=\sum_{n=1}^{\infty}\frac{F_{n}(R)}{R^{2}}\left[\sum_{i=1}^{3}\Phi_{n}^{(i)}(\alpha_{i};R)\right]\label{eq:WFhyperspherical}
\end{equation}
where at large distances $R\gg\Lambda^{-1}$ the hyper-radial functions
$F_{n}(R)$ are solutions of
\begin{equation}
\left(-\partial_{R}^{2}+\frac{s_{n}^{2}(R)-1/4}{R^{2}}-\frac{2mE}{\hbar^{2}}\right)\sqrt{R}F_{n}(R)=0\label{eq:Hyper-radial-equations}
\end{equation}
and the hyper-angular wave functions $\Phi_{n}^{(i)}$ are given
by~\cite{Braaten2006,Naidon2017} 
\begin{equation}
\Phi_{n}^{(i)}(\alpha;R)=\lambda^{(i)}\frac{\sin\left[s_{n}(R)\left(\frac{\pi}{2}-\alpha\right)\right]}{\sin2\alpha}\label{eq:hyperangularPhi}
\end{equation}
where $s_{n}(R)$ are the solutions of 
\begin{align}
\left(-\cos\left(s\frac{\pi}{2}\right)+\frac{2}{s}\frac{\sin(s\gamma)}{\sin2\gamma}+\frac{\sin\left(s\frac{\pi}{2}\right)}{s}\frac{R}{a_{23}}\right)\times\qquad\nonumber \\
\left(-\cos\left(s\frac{\pi}{2}\right)+\frac{\sin\left(s\frac{\pi}{2}\right)}{s}\frac{R}{a_{12}}\right)=2\left(\frac{2\sin(s\gamma^{\prime})}{s\sin2\gamma^{\prime}}\right)^{2}\label{eq:snEquation}
\end{align}
with $\gamma=\arcsin\frac{1}{1+A}$ and $\gamma^{\prime}=\frac{\pi}{4}-\frac{\gamma}{2}$.

\begin{figure}
\includegraphics[width=8cm]{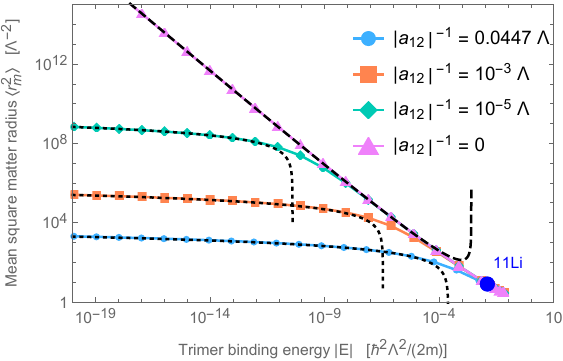}

\caption{\label{fig:MatterRadius}Mean square matter radius of the ground-state
trimer as a function of its binding energy $\vert E\vert$, obtained
numerically from the separable model. The different curves correspond
to different values of the particle-particle scattering length $a_{12}$
shown by the vertical lines in Fig.~\ref{fig:Close-up-figure}. The
dashed curve is obtained from the analytic formula Eq.~(\ref{eq:LogarithmicDivergence})
with $R_{1}$ set to $20\Lambda^{-1}$, and the dotted curves are
obtained from the analytic formula Eq.~(\ref{eq:QuadraticDivergence})
with $(R_{1},R_{2})$ set respectively to $(2.6,16)10^{4}$, $(420,1700)$,
$(34,64)$ in units of $\Lambda^{-1}$.}

\end{figure}

For $1/\vert a_{12}\vert\ne0$, the solutions $s_{n}(R)\xrightarrow[R\to\infty]{}2n$.
In this case, the trimer's extent diverges logarithmically with vanishing
binding energy in the channel $n=1$, while it remains finite in other
channels, as discussed long ago in Ref.~\cite{Fedorov1994a}. As
a result, the wave function is dominated by the channel $n=1$ at
large $R$, and from $s_{1}(R)\to2$ one can check from Eq.~(\ref{eq:hyperangularPhi})
that it does not depend on the hyper-angles. It follows that $\langle y_{1}^{2}\rangle=\langle y_{3}^{2}\rangle$~\cite{Fedorov1994a},
thus $\langle R_{23,1}^{2}\rangle/\langle R_{12,3}^{2}\rangle=\left(\mu_{23,1}/\mu_{12,3}\right)^{1/2}$
, and from Eq.~(\ref{eq:SquareRadiusRatio}) one recovers the limit
$\langle r_{m}^{2}\rangle/\langle r_{c}^{2}\rangle=A$ seen in the
right part of Fig.~\ref{fig:Ratio}. The logarithmic divergence
of the mean square radii in this limit is illustrated by the three
lowest curves (circles, squares, and diamonds) in Fig.~\ref{fig:MatterRadius}
corresponding to finite values of $a_{12}$.

The same hyper-spherical analysis can be carried out at the two particles'
resonance $1/\vert a_{12}\vert=0$. In that case, the solutions $s_{n}(R)\xrightarrow[R\to\infty]{}n$.
As a result, the channel $n=1$ is again dominant at large hyper-radius,
but it now leads to a divergence of the trimer's extent that is slightly
slower than the inverse of the energy. This divergence is illustrated
by the top curve (triangles) in Fig\@.~\ref{fig:MatterRadius} corresponding
to an infinite $a_{12}$. In this limit, the hyper-radius and hyper-angles
separate again, but there is now a dependence on the hyper-angles.
From the hyper-angular functions $\Phi_{1}^{(i)}$, one can recover\footnote{See Appendix \ref{sec:Hyper-spherical} for details.}
the ratio $\langle r_{m}^{2}\rangle/\langle r_{c}^{2}\rangle=\frac{2}{3}A$
seen in the left part of Fig.~\ref{fig:Ratio}.

It is important to note that the above results rely only on the behaviour
of $s_{n}^{2}(R)$ at large hyper-radius $R\gg\vert a_{12}\vert,\vert a_{23}\vert$,
where the hyper-radial potential $[s_{n}^{2}(R)-1/4]/R^{2}$ of Eq.\ (\ref{eq:Hyper-radial-equations})
is repulsive. They are not related to the Efimov effect, i.e. the
appearance of an attractive part (negative value of $s_{n}^{2}(R)=-\vert s_{0}\vert^{2}$)
in the hyper-radial potential at shorter hyper-radius $R\ll\vert a_{12}\vert,\vert a_{23}\vert$.
As a matter of fact, none of the analytical results presented here
depend on the quantity $s_{0}$ characterising Efimov universality.

In addition, the above results apply to any Borromean halo state close
to three-body dissociation. Thus, in a system where the Efimov effect
occurs, they apply equally to the ground state and all Efimov states.
In other words, the ground state and excited states have exactly the
same geometry close to their dissociation threshold. This halo universality
is explained by the fact that the hyper-angular part of the wave function
is the same for all states close to their threshold, giving a universal
shape distribution to all of these states. In contrast, the Efimov
universality, which is the discrete scale invariance of the spectrum
near the unitarity point, applies only approximately to the ground
state because the hyper-radial part of its wave function significantly
differs from the rescaled hyper-radial wave function of higher states,
thus deviating from the discrete scaling invariance.
\begin{figure*}[t]
\includegraphics[height=8cm]{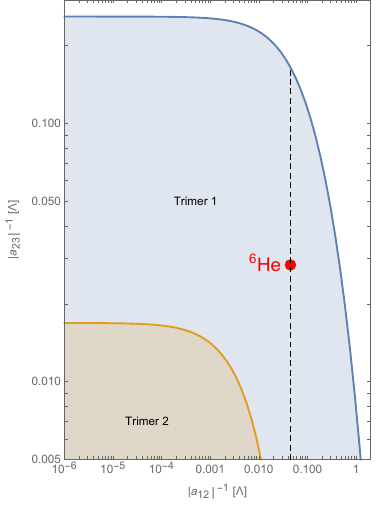}\includegraphics[viewport=15bp 0bp 181bp 243bp,clip,height=8cm]{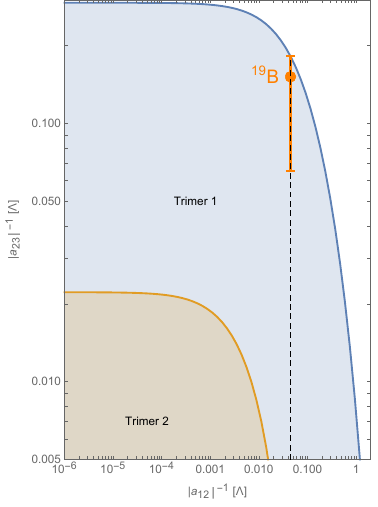}\includegraphics[viewport=15bp 0bp 181bp 243bp,clip,height=8cm]{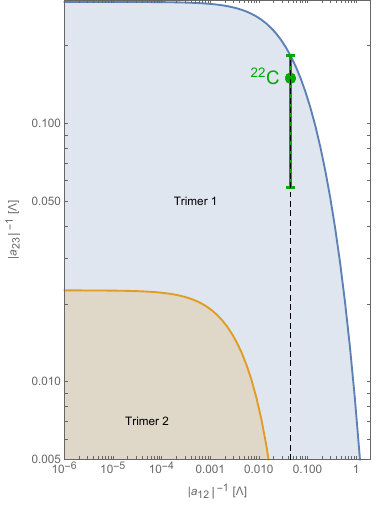}

\caption{\label{fig:Mass-ratio-4-17-20}Same figure as Fig.~\ref{fig:Close-up-figure}
for mass ratio 4 (left), 17 (middle) and 20 (right). The points corresponding
to helium-6, boron-19, and carbon-22 are obtained using the two-neutron
separation energies $\vert E({}^{6}\text{He})\vert=975.45\pm0.05$
keV, $\vert E({}^{19}\text{B})\vert=90\pm560$ keV, and $\vert E({}^{22}\text{C})\vert=100\pm640$
keV from AME2016~\cite{Wang2017}.}
\end{figure*}

\begin{figure}

\includegraphics[width=8cm]{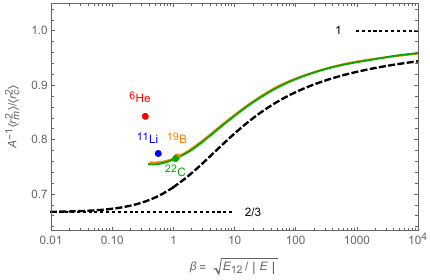}\caption{\label{fig:HaloNuclei}Matter/core radius ratio $\langle r_{m}^{2}\rangle/\langle r_{c}^{2}\rangle$
normalised by $A$ and energy ratio $\beta=\sqrt{E_{12}/\vert E\vert}$
for different halo nuclei. The solid curves indicate the possible
values for boron-19 and carbon-22 due to the current uncertainty on
their energy $E$, with $E_{12}$ being fixed by the neutron-neutron
scattering length $a_{12}=-(0.0447\Lambda)^{-1}$ -- this value is
indicated by the vertical dashed lines in Fig.~\ref{fig:Mass-ratio-4-17-20}.
The dashed curve corresponds to the analytical formula given by Eqs.~~(\ref{eq:SquareRadiusRatioTer},~\ref{eq:f1explicit},~\ref{eq:f2explicit}). }
\end{figure}

\section{Observing halo universality\label{sec:Observation}}

Two-neutron halo nuclei are prime candidates for the experimental
evidence of halo universality, since neutrons are nearly resonant,
and mean square radii can be extracted from experiments. Although
the separable potential model is not a precise description of these
halo nuclei, it can be used to estimate their universal nature. The
case of the lithium\nobreakdash-11 halo nucleus is represented as
a blue point in Figs.\ \ref{fig:Close-up-figure},~\ref{fig:Ratio},
and~\ref{fig:MatterRadius}. This point is obtained by setting $a_{12}^{-1}=-0.0447\Lambda$
and $\Lambda^{-1}=0.84$ fm to reproduce the neutron-neutron scattering
length $a_{12}=-18.8$\,fm and effective range $r_{12}=2.83\thinspace$fm
of the AV18 model\ \cite{Wiringa1995}, and setting $a_{23}^{-1}=-0.344\vert a_{23}^{(1)}\vert^{-1}$
to reproduce the two-neutron separation energy $\vert E\vert=369.15(65)$\,keV
of lithium\nobreakdash-11\ \cite{Smith2008}. These parameters lead
to a mean-square radius $\langle R_{12,3}^{2}\rangle=4.70$\,fm that
is in reasonable agreement with the value 5.01$\pm0.32$\,fm\ \cite{Shulgina2009}
derived from experimental data. One can see from Fig.~\ref{fig:Ratio}
that $^{11}$Li is not accurately described by the analytical limit
Eq.\,(\ref{eq:SquareRadiusRatioTer}), since its ratio $\langle r_{m}^{2}\rangle/\langle r_{c}^{2}\rangle$
is around 7.0 whereas Eq.\,(\ref{eq:SquareRadiusRatioTer}) gives
6.3, i.e. an error of 10\%. Also note that $^{11}$Li is yet too strongly
bound to enter the regime of logarithmic divergence of its size, as
shown in Fig.~\ref{fig:MatterRadius}.

The situation of other two-neutron halo nuclei is illustrated in Fig.~\ref{fig:Mass-ratio-4-17-20}.
In all cases, the core-neutron scattering length $a_{23}$ is set
to reproduce the two-neutron separation energies compiled by AEM2016~\cite{Wang2017}.
The matter/core radius ratio $\langle r_{m}^{2}\rangle/\langle r_{c}^{2}\rangle$
of these halo nuclei is shown in Fig.~\ref{fig:HaloNuclei}. It appears
that none of them lie in the regime where the analytical result of
Eqs.~(\ref{eq:SquareRadiusRatioTer},~\ref{eq:f1explicit},~\ref{eq:f2explicit})
is accurate. A prominent reason is that the neutron-neutron scattering
length $a_{12}$ is yet too small to fully reach this regime: it would
need to be significantly larger to allow $a_{23}$ to approach the
critical value $a_{23}^{(1)}$. Nevertheless, the cases of boron-19
and carbon-22 are the closest examples to halo universality, since
their matter/core radius ratio is possibly less than a few percent
off from the analytical formula. Further experimental determination
of their binding energy and geometric properties could confirm this
situation.

Ultracold mixtures of light and heavy atoms constitute another promising
platform for the observation of halo universality. The advantage of
such systems over atomic nuclei is that the scattering length between
the light particles may be controlled by a magnetic Fano-Feshbach
resonance. Examples include mixtures of caesium-133 and lithium-7
atoms, or lithium-6 atoms in different hyperfine states. However,
experiments with these systems are known to be hindered by strong
losses. In addition, specific experimental techniques such as Coulomb
explosion imaging~\cite{Voigtsberger2014} should be implemented
to measure geometrical properties such as the mean-square radii of
ultracold atomic trimers.

\section{Conclusion\label{sec:Conclusion}}

In summary, this work presents a general picture of the geometric
properties of trimer halos formed by a core and two particles, extending
previous findings to the whole spectrum of three-body bound states,
including Efimov states. It appears that close to the three-body dissociation
threshold of any state, the trimer forms an extended halo with universal
geometric properties that are given by the universal laws described
in Ref.~\cite{Fedorov1994a} away from the particle's resonance,
and by the laws recently found in Ref.~\cite{Hongo2022} close to
the particles' resonance. Both limits can be understood in the hyper-spherical
picture, which shows that the shape of the trimer is independent of
its size at the three-body dissociation threshold. In the first (off-resonant)
limit, the trimer's size increases logarithmically with vanishing
trimer binding energy, while it increases almost as the inverse of
the binding energy in the second (resonant) limit. This halo universality
is independent of the Efimov effect, and thus applies indistinctly
to all states, including Efimov states. It may be evidenced experimentally
in two-neutron halo nuclei or ultracold atomic mixtures. Upon completion
of this work, a related study~\cite{Pang2023} reported an extension
of the formula Eq.~(\ref{eq:SquareRadiusRatioTer}) using effective-field
theory techniques.

\noindent %
\noindent\begin{minipage}[t]{1\columnwidth}%
\rule[0.5ex]{1\columnwidth}{1pt}%
\end{minipage}

\subsection*{Acknowledgments}

The author acknowledges support from the JSPS Grants-in-Aid for Scientific
Research on Innovative Areas (No. JP18H05407). He is thankful to L.~Happ,
M.~Hongo, L.~Pricoupenko, and J.~Dalibard for helpful discussions.
\hypersetup{urlcolor=myred}

\noindent {\small{}{} \bibliographystyle{IEEEtran2}
\bibliography{paper39}
 }{\small\par}

\clearpage{}

\appendix
\begin{strip}\textsf{\textbf{\huge{}Appendix}}\end{strip}

\section{Low-momentum and energy T-matrix element Eq.~(\ref{eq:Off-shell-value})\label{sec:Low-momentum-and-energy}}

The Lippman-Schwinger Eq.~(\ref{eq:TwoBodyTransitionOperator}) defining
the two-body transition operator can be written explicitly (dropping
here the indices $ij$),

\begin{equation}
\langle\bm{k}\vert\hat{T}(z)\vert\bm{q}\rangle=\langle\bm{k}\vert\hat{V}\vert\bm{q}\rangle+\int\frac{d^{3}\bm{k}^{\prime}}{(2\pi)^{3}}\langle\bm{k}\vert\hat{V}\vert\bm{k}^{\prime}\rangle\frac{\langle\bm{k}^{\prime}\vert T(z)\vert\bm{q}\rangle}{z^{+}-\frac{\hbar^{2}k^{\prime2}}{2\mu}}.\label{eq:Lippman-Schwinger}
\end{equation}
with $z^{+}\equiv z+i\epsilon$. The interaction $\hat{V}$ (and thus
$\hat{T}$) having a finite range $\Lambda^{-1}$, one can set $\bm{k}\approx\bm{0}$
and $\bm{q}\approx\bm{0}$ in the matrix elements for $k,q\ll\Lambda$
within an error of order $O(k^{2},q^{2})$. This gives

\begin{equation}
\frac{1}{\langle\bm{0}\vert\hat{T}(z)\vert\bm{0}\rangle}=\frac{1}{\langle\bm{0}\vert\hat{V}\vert\bm{0}\rangle}+\int\frac{d^{3}\bm{k}}{(2\pi)^{3}}\Upsilon(\bm{k},z)\frac{1}{\frac{\hbar^{2}k^{2}}{2\mu}-z^{+}}\label{eq:LowMomenta}
\end{equation}
with the function
\begin{equation}
\Upsilon(\bm{k},z)\equiv\frac{\langle\bm{0}\vert\hat{V}\vert\bm{k}\rangle}{\langle\bm{0}\vert\hat{V}\vert\bm{0}\rangle}\frac{\langle\bm{k}\vert T(z)\vert\bm{0}\rangle}{\langle\bm{0}\vert\hat{T}(z)\vert\bm{0}\rangle}\xrightarrow[k\ll\Lambda]{}1.\label{eq:Definition_of_f}
\end{equation}
Now, adding a counter-term in the integral, one can write:
\begin{align}
\frac{1}{\langle\bm{0}\vert\hat{T}(z)\vert\bm{0}\rangle} & =\frac{1}{\langle\bm{0}\vert\hat{V}\vert\bm{0}\rangle}\nonumber \\
 & +\int\frac{d^{3}\bm{k}}{(2\pi)^{3}}\Upsilon(\bm{k},z)\left(\frac{1}{\frac{\hbar^{2}k^{2}}{2\mu}-z^{+}}-\frac{1}{\frac{\hbar^{2}k^{2}}{2\mu}}\right)\nonumber \\
 & +\int\frac{d^{3}\bm{k}}{(2\pi)^{3}}\Upsilon(\bm{k},z)\frac{1}{\frac{\hbar^{2}k^{2}}{2\mu}}\label{eq:LowMomenta2}
\end{align}
Taking the limit $z\to0$ gives
\begin{equation}
\frac{1}{\langle\bm{0}\vert\hat{T}(0)\vert\bm{0}\rangle}=\frac{1}{\langle\bm{0}\vert\hat{V}\vert\bm{0}\rangle}+\int\frac{d^{3}\bm{k}}{(2\pi)^{3}}\Upsilon(\bm{k},0)\frac{1}{\frac{\hbar^{2}k^{2}}{2\mu}},\label{eq:ZeroEnergyLimit}
\end{equation}
so that one can eliminate $\langle\bm{0}\vert\hat{V}\vert\bm{0}\rangle$
in favour of $\langle\bm{0}\vert\hat{T}(0)\vert\bm{0}\rangle$,
\begin{align}
\frac{1}{\langle\bm{0}\vert\hat{T}(z)\vert\bm{0}\rangle} & =\frac{1}{\langle\bm{0}\vert\hat{T}(0)\vert\bm{0}\rangle}\nonumber \\
 & +\int\frac{d^{3}\bm{k}}{(2\pi)^{3}}\Upsilon(\bm{k},z)\left(\frac{1}{\frac{\hbar^{2}k^{2}}{2\mu}-z^{+}}-\frac{1}{\frac{\hbar^{2}k^{2}}{2\mu}}\right)\nonumber \\
 & +\int\frac{d^{3}\bm{k}}{(2\pi)^{3}}\left(\Upsilon(\bm{k},z)-\Upsilon(\bm{k},0)\right)\frac{1}{\frac{\hbar^{2}k^{2}}{2\mu}}\label{eq:LowMomenta3}
\end{align}
Finally, one can consider the low-energy limit $z\ll\hbar^{2}\Lambda^{2}/2\mu$.
In this limit, by virtue of Eq.~(\ref{eq:Definition_of_f}), the
second line of Eq.~(\ref{eq:LowMomenta3}) can be approximated as
\[
\underbrace{\Upsilon(\bm{0},z)}_{1}\frac{2\mu}{\hbar^{2}}\int\frac{d^{3}\bm{k}}{(2\pi)^{3}}\left(\frac{1}{k^{2}-\frac{2\mu z^{+}}{\hbar^{2}}}-\frac{1}{k^{2}}\right)=\frac{2\mu}{4\pi\hbar^{2}}i\sqrt{\frac{2\mu z}{\hbar^{2}}}
\]
while the third line of Eq.~(\ref{eq:LowMomenta3}) may be neglected,
assuming that $\Upsilon(\bm{k},z)-\Upsilon(\bm{k},0)\sim O(z)$. Using
the standard result $\langle\bm{0}\vert\hat{T}(0)\vert\bm{0}\rangle=4\pi\hbar^{2}a/(2\mu)$
where $a$ is the scattering length, one obtains Eq.~(\ref{eq:Off-shell-value}),
which is valid with an error of order $O(k^{2},k^{\prime2},z_{ij})$.

\section{Functions $f_{n}$ anf $f_{c}$ \label{sec:Functions}}

\renewcommand{\theequation}{\thesection.\arabic{equation}}
\setcounter{equation}{0}

\subsection{Numerator of Eq.~(\ref{eq:SquareRadius32-1})}

Retaining only the Faddeev component $\mathcal{F}_{12}\equiv\mathcal{F}$
associated with the two particles in Eq.~(\ref{eq:ThreeBodyWavefunction}),
the integral in the numerator of Eq.~(\ref{eq:SquareRadius32-1})
is expressed as
\begin{equation}
\int d^{3}K_{23,1}d^{3}k_{23}\left|\bm{\nabla}_{\bm{K}_{23,1}}\frac{\mathcal{F}(\vert\bm{k}_{23}-\frac{A}{A+1}\bm{K}_{23,1}\vert)}{\frac{\hbar^{2}}{2\mu_{23,1}}K_{23,1}^{2}+\frac{\hbar^{2}}{2\mu_{23}}k_{23}^{2}-E}\right|^{2}\label{eq:Numerator23-1A}
\end{equation}
where we used $\bm{K}_{12,3}=\bm{k}_{23}-\frac{A}{A+1}\bm{K}_{23,1}$.
Using the relation
\begin{equation}
\nabla_{\bm{p}}\mathcal{F}(\vert\alpha\bm{p}+\beta\bm{q}\vert)=\alpha\frac{\alpha\bm{p}+\beta\bm{q}}{\vert\alpha\bm{p}+\beta\bm{q}\vert}\mathcal{F}^{\prime}(\vert\alpha\bm{p}+\beta\bm{q}\vert)\label{eq:Gradient}
\end{equation}
and re-expressing the integrand in terms of $\bm{k}\equiv\bm{k}_{12}$
and $\bm{K}\equiv\bm{K}_{12,3}$ by using $\bm{K}_{23,1}=-\bm{k}_{12}-\frac{1}{2}\bm{K}_{12,3}$,
one arrives at
\begin{multline}
\int d^{3}Kd^{3}k\Bigg\vert-\left(\frac{A}{A+1}\right)\frac{\frac{\bm{K}}{K}\mathcal{F}^{\prime}(K)}{\frac{\hbar^{2}}{2\mu_{12,3}}K^{2}+\frac{\hbar^{2}}{2\mu_{12}}k^{2}-E}\\
-\frac{\frac{\hbar^{2}}{2\mu_{23,1}}2\left(-\bm{k}-\frac{1}{2}\bm{K}\right)\mathcal{F}(K)}{\left(\frac{\hbar^{2}}{2\mu_{12,3}}K^{2}+\frac{\hbar^{2}}{2\mu_{12}}k^{2}-E\right)^{2}}\Bigg\vert^{2}\label{eq:Numerator23-1B}
\end{multline}
Expanding the square, the cross term proportional to $\bm{K}\cdot\bm{k}$
averages to zero since the orientations of $\bm{K}$ and $\bm{k}$
are independent. Integrating the remaining terms over $\bm{k}$ by
using
\begin{align}
\int_{0}^{\infty}k^{2}dk\frac{1}{\left(k^{2}+Q^{2}\right)^{2}} & =\frac{\pi}{4Q}\label{eq:Integral1}\\
\int_{0}^{\infty}k^{2}dk\frac{1}{\left(k^{2}+Q^{2}\right)^{3}} & =\frac{\pi}{16Q^{3}}\label{eq:Integral2}\\
\int_{0}^{\infty}k^{2}dk\frac{1}{\left(k^{2}+Q^{2}\right)^{4}} & =\frac{\pi}{32Q^{5}}\label{eq:Integral3}\\
\int_{0}^{\infty}k^{2}dk\frac{k^{2}}{\left(k^{2}+Q^{2}\right)^{4}} & =\frac{\pi}{32Q^{3}}\label{eq:Integral4}
\end{align}
one arrives at
\begin{equation}
\boxed{\left(\pi\frac{2\mu_{12}}{\hbar^{2}}\right)^{2}\sqrt{\frac{\mu_{12,3}}{\mu_{12}}}\left[\left(\frac{A}{A+1}\right)^{2}f_{c}+\frac{A\left(A+2\right)}{A+1}f_{n}\right]}\label{eq:Numerator23-1C}
\end{equation}
with
\begin{align}
f_{c} & \equiv\int d^{3}K\left(\frac{\mathcal{F}^{\prime}(K)^{2}}{\tilde{K}}-\frac{K\mathcal{F}^{\prime}(K)\mathcal{F}(K)}{\tilde{K}^{3}}+\frac{K^{2}\mathcal{F}(K)^{2}}{2\tilde{K}^{5}}\right)\label{eq:Definitionfc}\\
f_{n} & \equiv\frac{1}{2}\int d^{3}K\frac{\mathcal{F}(K)^{2}}{\tilde{K}^{3}}\label{eq:Definitionfn}
\end{align}
and $\tilde{K}^{2}\equiv K^{2}-\frac{2\mu_{12,3}E}{\hbar^{2}}$.

\subsection{Numerator of Eq.~(\ref{eq:SquareRadius12-3})}

Retaining only the Faddeev component $\mathcal{F}_{12}\equiv\mathcal{F}$
associated with the two particles in Eq.~(\ref{eq:ThreeBodyWavefunction}),
the integral in the numerator of Eq.~(\ref{eq:SquareRadius12-3})
is expressed as
\begin{equation}
\int d^{3}Kd^{3}k\left|\bm{\nabla}_{\bm{K}}\frac{\mathcal{F}(\bm{K})}{\frac{\hbar^{2}}{2\mu_{12,3}}K^{2}+\frac{\hbar^{2}}{2\mu_{12}}k^{2}-E}\right|^{2}\label{eq:Numerator12-3A}
\end{equation}
where $\bm{K}=\bm{K}_{12,3}$ and $\bm{k}=\bm{k}_{12}$. This gives
\begin{multline}
\int d^{3}Kd^{3}k\Bigg(\frac{\mathcal{F}^{\prime}(K)}{\frac{\hbar^{2}}{2\mu_{12,3}}K^{2}+\frac{\hbar^{2}}{2\mu_{12}}k^{2}-E}\\
-\frac{\frac{\hbar^{2}}{2\mu_{12,3}}2K\mathcal{F}(K)}{\left(\frac{\hbar^{2}}{2\mu_{12,3}}K^{2}+\frac{\hbar^{2}}{2\mu_{12}}k^{2}-E\right)^{2}}\Bigg)^{2}\label{eq:Numerator12-3B}
\end{multline}
Expanding the square and integrating over $\bm{k}$, one arrives at:
\begin{equation}
\boxed{\left(\pi\frac{2\mu_{12}}{\hbar^{2}}\right)^{2}\sqrt{\frac{\mu_{12,3}}{\mu_{12}}}f_{c}}\label{eq:Numerator12-3C}
\end{equation}
where we used Eq.~(\ref{eq:Integral1}-\ref{eq:Integral3}).

\subsection{Ratio of mean square radii}

Using Eqs.~(\ref{eq:Numerator23-1C}) and (\ref{eq:Numerator12-3C})
in the expression of the ratio of mean square radii Eq.~(\ref{eq:SquareRadiusRatio}),
one obtains
\begin{align*}
\frac{\langle r_{m}^{2}\rangle}{\langle r_{c}^{2}\rangle} & =\frac{\frac{1}{2}\left(A+1\right)^{2}\frac{\langle R_{23,1}^{2}\rangle}{\langle R_{12,3}^{2}\rangle}+A}{A+2}\\
 & =\frac{1}{2}A\left(1+\frac{f_{n}}{f_{c}}\right)
\end{align*}
which yields Eq.~(\ref{eq:SquareRadiusRatioTer}) of the main text.

\subsection{Simplification of $f_{n}$ and $f_{c}$}

The functions $f_{n}$ and $f_{c}$ given by Eqs.~(\ref{eq:Definitionfc}-\ref{eq:Definitionfn})
can be expressed with the dimensionless variable $q\equiv\sqrt{\frac{\hbar^{2}}{2\mu_{12,3}\vert E\vert}}K$,
\begin{align}
f_{n} & (\beta)\propto\frac{1}{2}\int d^{3}q\frac{\mathcal{F}(q)^{2}}{\tilde{q}^{3}}\label{eq:fnq}\\
f_{c} & (\beta)\propto\int d^{3}q\left(\frac{\mathcal{F}^{\prime}(q)^{2}}{\tilde{q}}-\frac{q\mathcal{F}^{\prime}(q)\mathcal{F}(q)}{\tilde{q}^{3}}+\frac{q^{2}\mathcal{F}(q)^{2}}{2\tilde{q}^{5}}\right)\label{eq:fcq}
\end{align}
where $\tilde{q}^{2}\equiv q^{2}+1$, and 
\begin{equation}
\mathcal{F}(q)\propto\frac{1}{\beta+\tilde{q}}.\label{eq:Fq}
\end{equation}
The function $f_{c}$ can be simplified as
\begin{align}
f_{n}(\beta) & \propto2\pi\int_{0}^{\infty}\frac{q^{2}dq}{\tilde{q}^{3}\left(\beta+\tilde{q}\right)^{2}}\label{eq:fnqSimple}\\
f_{c}(\beta) & \propto2\pi\int_{0}^{\infty}\frac{dq}{\tilde{q}\left(\beta+\tilde{q}\right)^{2}}\label{eq:fcqSimple}
\end{align}

The last expression can be verified by successive integrations by
parts of Eq.~(\ref{eq:fnqSimple}), $\int u^{\prime}v=-\int uv^{\prime}$,
with first $u(q)=q$, $v(q)=\tilde{q}^{-1}\left(\beta+\tilde{q}\right)^{-2}$,
and then $u(q)=q^{3}/3;v(q)=\tilde{q}^{-3}\left(\beta+\tilde{q}\right)^{-2}+2\tilde{q}^{-2}\left(\beta+\tilde{q}\right)^{-3}$,
leading to the original form of $f_{c}$ in Eq.~(\ref{eq:fcq}).
Finally, making the change of variable $y\equiv q^{2}+1$, one arrives
at Eqs.~(\ref{eq:f1simplified}-\ref{eq:f2simplified}) of the main
text.

\section{Hyper-spherical representation\label{sec:Hyper-spherical}}

In the hyper-spherical representation, the values of the scattering
lengths $a_{23}=a_{31}$ and $a_{12}$ for the corresponding pairs
can be imposed in the wave function Eq.~(\ref{eq:WFhyperspherical})
by applying Bethe-Peierls conditions~\cite{Braaten2006,Naidon2017}.
This leads to the following equations on the coefficients $\lambda^{(1)}=\lambda^{(2)}$
and $\lambda^{(3)}$: 
\begin{align}
M_{11}\lambda^{(1)}+M_{13}\lambda^{(3)} & =0\label{eq:EquationsOnlambdas}\\
M_{31}\lambda^{(1)}+M_{33}\lambda^{(3)} & =0\nonumber 
\end{align}
with 
\begin{align}
M_{11} & =\left(-s\cos\left(s\frac{\pi}{2}\right)+\frac{R}{a_{23}}\sin\left(s\frac{\pi}{2}\right)\right)+2\frac{\sin s\gamma}{\sin2\gamma}\label{eq:M11}\\
M_{13} & =2\frac{\sin s\gamma^{\prime}}{\sin2\gamma^{\prime}}\label{eq:M12}\\
M_{31} & =4\frac{\sin s\gamma^{\prime}}{\sin2\gamma^{\prime}}\label{eq:M21}\\
M_{33} & =-s\cos\left(s\frac{\pi}{2}\right)+\frac{R}{a_{12}}\sin\left(s\frac{\pi}{2}\right)\label{eq:M22}
\end{align}
To obtain non-zero solutions $\lambda^{(1)}$ and $\lambda^{(3)}$,
the determinant of Eq.~(\ref{eq:EquationsOnlambdas}) should be zero,
which leads to
\begin{align}
\left(-\cos\left(s\frac{\pi}{2}\right)+\frac{2}{s}\frac{\sin(s\gamma)}{\sin2\gamma}+\frac{\sin\left(s\frac{\pi}{2}\right)}{s}\frac{R}{a_{23}}\right)\times\qquad\label{eq:snEquation2}\\
\left(-\cos\left(s\frac{\pi}{2}\right)+\frac{\sin\left(s\frac{\pi}{2}\right)}{s}\frac{R}{a_{12}}\right)=2\left(\frac{2\sin(s\gamma^{\prime})}{s\sin2\gamma^{\prime}}\right)^{2}\nonumber 
\end{align}
which is Eq.~(\ref{eq:snEquation}) of the main text. 

\subsection{Mean-square radii for vanishing binding energy}

In the hyper-spherical representation, any mean square radius is given
by an expression of the form,
\begin{equation}
\langle r^{2}\rangle=\frac{\sum_{n,n^{\prime}}\cdots\int dR\,R^{2}\left(\sqrt{R}F_{n}(R)\right)\left(\sqrt{R}F_{n^{\prime}}(R)\right)}{\sum_{n,n^{\prime}}\cdots\int dR\,\left(\sqrt{R}F_{n}(R)\right)\left(\sqrt{R}F_{n^{\prime}}(R)\right)}.\label{eq:GenericMeanSquareRadius}
\end{equation}
where the dots indicate numerical factors resulting from integration
over the hyper-angles. From Eq.\ (\ref{eq:Hyper-radial-equations})
of the main text, at sufficiently large hyper-radius $R\gg R_{0}$
(to be specified below), the hyper-radial functions $F_{n}(R)$ satisfy
the following equation:
\begin{equation}
\left(-\partial_{R}^{2}+\frac{s_{n}^{2}(\infty)-1/4}{R^{2}}-\frac{2mE}{\hbar^{2}}\right)\sqrt{R}F_{n}(R)=0\label{eq:Hyper-radialEquation}
\end{equation}
where $s_{n}(\infty)=\lim_{R\to\infty}s_{n}(R)$. It follows that
\begin{equation}
F_{n}(R)\xrightarrow[R\gg R_{0}]{}K_{s_{n}(\infty)}(\kappa R)\label{eq:AsymptoteOfFn}
\end{equation}
 where $\kappa=-2mE/\hbar^{2}$, and $K$ designates the modified
Bessel function of the second kind. 

\subsubsection*{Off-resonance case $1/\vert a_{12}\vert\protect\ne0$}

The solutions $s_{n}$ of Eq.~(\ref{eq:snEquation2}) behave as $s_{n}(R)\xrightarrow[R\gg R_{0}]{}2n$,
with $R_{0}\sim\max\left(\vert a_{12}\vert,\vert a_{23}\vert\right)$.
In this case, only the lowest channel $n=1$ gives a hyper-radial
function that extends far beyong $R_{0}$ for vanishing energy $E$.
Close to the three-body dissociation threshold, one can thus neglect
the other channels in the calculation of the mean square radius Eq.~(\ref{eq:GenericMeanSquareRadius}),
and using Eq.~(\ref{eq:AsymptoteOfFn}) one finds
\begin{equation}
\boxed{\langle r^{2}\rangle\xrightarrow[\kappa\ll R_{0}^{-1}]{}\eta\langle R^{2}\rangle=\eta R_{1}^{2}\ln\frac{1}{\kappa R_{2}}},\label{eq:LogarithmicDivergence}
\end{equation}
where the distances $R_{1},R_{2}\gtrsim R_{0}$ are independent of
$\kappa$. This shows that the mean square radius diverges logarithmically
with vanishing trimer energy. The coefficient $\eta$ results from
the hyper-angular integration.

\subsubsection*{On-resonance case $1/\vert a_{12}\vert=0$}

The solutions $s_{n}$ of Eq.~(\ref{eq:snEquation2}) behave as
$s_{n}(R)\xrightarrow[R\gg R_{0}]{}n$, with $R_{0}\sim\vert a_{23}\vert$.
In this case, both the lowest channel $n=1$ and second channel $n=2$
give hyper-radial functions that extend far beyond $R_{0}$ with vanishing
energy. However, the channel $n=1$ extends to distances $\sim\kappa^{-1}$,
whereas the $n=2$ channel extends to distances $\sim R_{0}\ln\frac{1}{\kappa R_{0}}$.
Close to the three-body dissociation threshold, the mean square radius
is therefore dominated again by the lowest channel $n=1$, which gives
\begin{equation}
\boxed{\langle r^{2}\rangle\xrightarrow[\kappa\ll R_{0}^{-1}]{}\eta\langle R^{2}\rangle=\eta\frac{2}{3\kappa^{2}}\left(\ln\frac{1}{\kappa R_{1}}\right)^{-1}},\label{eq:QuadraticDivergence}
\end{equation}
where the distance $R_{1}\gtrsim R_{0}$ is independent of $\kappa$.
This shows that the mean square radius diverges slightly more slowly
than the inverse of the vanishing energy $E$. 

\subsection{Calculation of the matter and the core radii}

Let us now calculate the matter and core radius explicitly. The coefficients
$\eta$ for the matter and core radii follow from Eq~(\ref{eq:SquareMatterRadius}-\ref{eq:rC3})
of the main text:
\begin{align}
\eta_{m} & =2\frac{\left(A+1\right)\langle\cos^{2}\alpha_{1}\rangle+\langle\cos^{2}\alpha_{3}\rangle}{\left(A+2\right)^{2}}\label{eq:MatterRadiusCoefficient}\\
\eta_{c} & =\frac{2}{A(A+2)}\langle\cos^{2}\alpha_{3}\rangle\label{eq:CoreRadiusCoefficient}
\end{align}

\subsubsection*{Off-resonance case $1/\vert a_{12}\vert\protect\ne0$}

In the dominant hyper-angular channel, $s_{1}(R)\to2$, therefore
the corresponding component is independent of hyper-angles, such that
$\langle\cos^{2}\alpha_{i}\rangle=1/2$. Therefore,

\begin{align}
\eta_{m} & =\frac{1}{A+2}\label{eq:MatterRadiusCoefficientOff}\\
\eta_{c} & =\frac{1}{A(A+2)}\label{eq:CoreRadiusCoefficientOff}
\end{align}
It follows that 
\begin{equation}
\boxed{\frac{\langle r_{m}^{2}\rangle}{\langle r_{c}^{2}\rangle}=\frac{\eta_{m}\langle R^{2}\rangle}{\eta_{c}\langle R^{2}\rangle}=A}\label{eq:RatioOffResonance}
\end{equation}

\subsubsection*{On-resonance case $1/\vert a_{12}\vert=0$}

From Eq.~(\ref{eq:EquationsOnlambdas}), one has
\begin{equation}
\frac{\lambda^{(3)}}{\lambda^{(1)}}=-\frac{M_{11}}{M_{13}}\label{eq:lambda3-over-lambda1}
\end{equation}
and in the limit $R\to\infty$, for the dominant hyper-angular channel
$s_{1}(R)\to1$, one finds
\begin{equation}
\frac{\lambda^{(3)}}{\lambda^{(1)}}=-\frac{\frac{R}{a_{23}}+\frac{1}{\cos\gamma}}{\frac{1}{\cos\gamma^{\prime}}}\to\infty\label{eq:lambda3-over-lambda1bis}
\end{equation}
Thus, in the calculation of hyper-angular averages, one can only retain
the component 
\begin{equation}
\Phi^{(3)}(\alpha;R)\propto\frac{1}{\sin\alpha}\label{eq:Phi3}
\end{equation}
(It is the same approximation as retaining only the Faddeev component
$\mathcal{F}_{12}$ in the calculation of Sec.~\ref{sec:Two-particle-resonance}).
This gives,
\begin{align}
\langle\cos^{2}\alpha_{3}\rangle & =\frac{\int_{0}^{\pi/2}d\alpha_{3}\sin^{2}(2\alpha_{3})\int_{-1}^{1}du_{3}\left(\cos^{2}\alpha_{3}\right)\left|\Phi^{(3)}(\alpha_{3})\right|^{2}}{\int_{0}^{\pi/2}d\alpha_{3}\sin^{2}(2\alpha_{3})\int_{-1}^{1}du_{3}\left|\Phi^{(3)}(\alpha_{3})\right|^{2}}\nonumber \\
 & =\frac{\int_{0}^{\pi/2}d\alpha\cos^{4}\alpha}{\int_{0}^{\pi/2}d\alpha\cos^{2}\alpha}=\frac{3}{4}\label{eq:cos2alpha3}
\end{align}
and 
\begin{align}
\langle\cos^{2}\alpha_{1}\rangle & =\frac{\int_{0}^{\pi/2}d\alpha_{3}\sin^{2}(2\alpha_{3})\int_{-1}^{1}du_{3}\left(\cos^{2}\alpha_{1}\right)\left|\Phi^{(3)}(\alpha_{3})\right|^{2}}{\int_{0}^{\pi/2}d\alpha_{3}\sin^{2}(2\alpha_{3})\int_{-1}^{1}du_{3}\left|\Phi^{(3)}(\alpha_{3})\right|^{2}}\nonumber \\
 & =\frac{1}{4}\left(\frac{1+2A}{1+A}\right)\label{eq:cos2alpha1}
\end{align}
where we used the change of variables $\int_{-1}^{1}du_{3}=\frac{2}{\sin2\gamma^{\prime}}\int_{\vert\frac{\pi}{2}-\alpha_{3}-\gamma^{\prime}\vert}^{\frac{\pi}{2}-\vert\alpha_{3}-\gamma^{\prime}\vert}\frac{\sin2\alpha_{1}}{\sin2\alpha_{3}}d\alpha_{1}$
and the definitions of $\gamma^{\prime}$ and $\gamma$ in terms of
$A$. From Eqs.~(\ref{eq:MatterRadiusCoefficient}-\ref{eq:CoreRadiusCoefficient})
one finds
\begin{align}
\eta_{m} & =\frac{1}{A+2}\label{eq:MatterRadiusCoefficientOn}\\
\eta_{c} & =\frac{3}{2}\frac{1}{A(A+2)}\label{eq:CoreRadiusCoefficientOn}
\end{align}
It follows that
\begin{equation}
\boxed{\frac{\langle r_{m}^{2}\rangle}{\langle r_{c}^{2}\rangle}=\frac{\eta_{m}\langle R^{2}\rangle}{\eta_{c}\langle R^{2}\rangle}=\frac{2}{3}A}\label{eq:RatioOnResonance}
\end{equation}

\end{document}